\documentclass[aps,prl,reprint,groupedaddress,showpacs]{revtex4-1}

\usepackage{amsfonts}
\usepackage{amssymb}
\usepackage{amsmath}
\usepackage{hyperref}
\usepackage{fullpage}
\usepackage{graphicx}
\usepackage{color}

\definecolor{red}{rgb}{1,0,0}

\begin{document}

\title{Critical wave functions in disordered graphene}
\author{J.E. Barrios-Vargas}
\author{Gerardo G. Naumis}
\email{naumis@fisica.unam.mx}
\affiliation{Depto. de F\'{i}sica-Qu\'{i}mica, Instituto de F\'{i}sica, Universidad Nacional Aut\'onoma de 
M\'exico (UNAM). Apdo. Postal 20-364, 01000, M\'exico D.F., M\'exico.}
\date{\today}

\begin{abstract}
In  order to elucidate the presence of non-localized states in doped graphene, an 
scaling analysis of the wave function moments known as inverse participation ratios is 
performed. The model used is a tight-binding hamiltonian considering nearest and next-nearest 
neighbors with random substitutional  impurities. Our findings indicate the presence of  
non-normalizable wave functions that follow a critical (power-law) decay, which are  between  a 
metallic and insulating behavior. The power-law exponent distribution is robust  against 
the inclusion of next-nearest neighbors and on growing the system size.
\end{abstract}
\maketitle

Graphene is a  two-dimensional atomic crystal 
\cite{novoselov1} with the highest known charge carrier mobility \cite{novoselov2} and  
thermal conductivity \cite{balandin} at room temperature. Both properties set up graphene as a 
raw material for transistors, however, the `{\it graphenium inside}' era is quite 
far \cite{geim}. Keeping in mind the aim of design a transistor, the problem turns out to be how to 
alchemize it into a semiconductor.  An alternative is to dope graphene. This leads immediately 
to the question of having quantum percolation in two dimensions, which has been the subject of debate many years ago \cite{meir,grest}. In the literature, usually it is found that ``there is no true 
metallic behavior in two dimensions'' as a consequence of the fact that all eigenstates are 
localized even when the disorder is weak \cite{abrahams}. In graphene, there has been a debate 
about this point \cite{amini,schleede,amini2}. Recently, disordered graphene has been classified 
using symmetry arguments around the Dirac point \cite{ostrovsky}; this classification allows a 
minimal conductivity 
behavior. Experimentally, the minimal-conductivity was measured for potassium atoms onto the 
graphene surface \cite{chen,yan}. Furthermore, it has been found that a metal-transition can be 
observed when graphene is doped with H \cite{bostwick}. Also, using a non-interacting electron model enriched 
by first-principles calculation the metal-insulator transition is observed \cite{leconte2010,leconte2011}. 
In this Letter, we present numerical evidence that shows a very interesting scenario. We have 
characterized the 
probability distribution of the 
moments associated to the wave function using the inverse participation ratios. The scaling of this 
quantity is frequently used to discriminated when an eigenstate is extended or localized. In doped 
graphene, we found extended states which do not follow the usual exponential localization; instead, 
these are critical, i.e., the wave function decay spatially as a non-normalizable power-law.  
This behavior evidenced the multifractality of the wave function \cite{evers}. Notice that in the orignal 
development of the scaling theory, critical states were not considered \cite{abrahams}.

As a model we use the tight-binding hamiltonian,
\begin{align}\label{hamand}
\mathcal{H}=-t\sum_{\langle i,j \rangle} c_i^\dag c_j 
-t' \sum_{\langle\langle i,j \rangle\rangle} c_i^\dag c_j 
+\varepsilon \sum_{\ell} c_\ell^\dag c_\ell \,,
\end{align}
where the nearest neighbor, $t=2.79\,{\rm eV}$, and next-nearest neighbor (NNN), 
$t'=0.68\,{\rm eV}$, 
hoping parameters are included; these values have been taken from reference~\cite{rubio}. The 
impurity sites, $\ell$, have been distributed randomly in the lattice according with a fixed 
concentration, $C$; and $\varepsilon$ is the self-energy for an impurity.

In order to investigate localization, we introduce the inverse p-participation ratios 
(IPRs),
\begin{align}\label{iprs}
|| \Psi_{{\mathbf{k}}} ||_{2p}= \sum_i^N  |\Psi_{{\mathbf{k}}}(\mathbf{r}_i)|^{2p} \,,
\end{align}
where $\Psi_{{\mathbf{k}}}$ is the wave function associated to the eigenstate $\mathbf{k}$ 
with energy $E_{\mathbf{k}}$,
which solves the Schr\"odinger equation $\mathcal{H} \Psi_{{\mathbf{k}}} = E_{\mathbf{k}}
\Psi_{{\mathbf{k}}} $. The index $i$ belongs to the sum over sites, $N$ is the total number of sites 
and $p$ is an integer. 
When $p=1$, $|| \Psi_{{\mathbf{k}}} ||_{2}=1$ because of the normalization condition. If 
the wave function of the eigenstate follows the power-law, 
$\Psi(\mathbf{r})_{{\mathbf{k}}} \sim |\mathbf{r}|^{-\alpha}$, the p-IPRs are scaled as~\cite{tsunetsugu},
\begin{align}
|| \Psi_{{\mathbf{k}}} ||_{2p} \simeq 
\left\{
\begin{array}{ll}
N^{-(p-1)} & (0\leq \alpha < \frac{1}{p}) \\
N^{-p(1-\alpha)} & (\frac{1}{p} \leq \alpha < 1) \,, \\
N^0 & (1\leq \alpha)
\end{array}
\right.
\end{align}
when $p>1$. The behavior $N^{-1}$ 
corresponds to a metal while $N^{0}$ corresponds to an insulator. Notice that here 
$N\propto L^2$, where $L$ it is the length sample. 

\begin{figure}[h]
\centering
\includegraphics[width=1.0\linewidth]{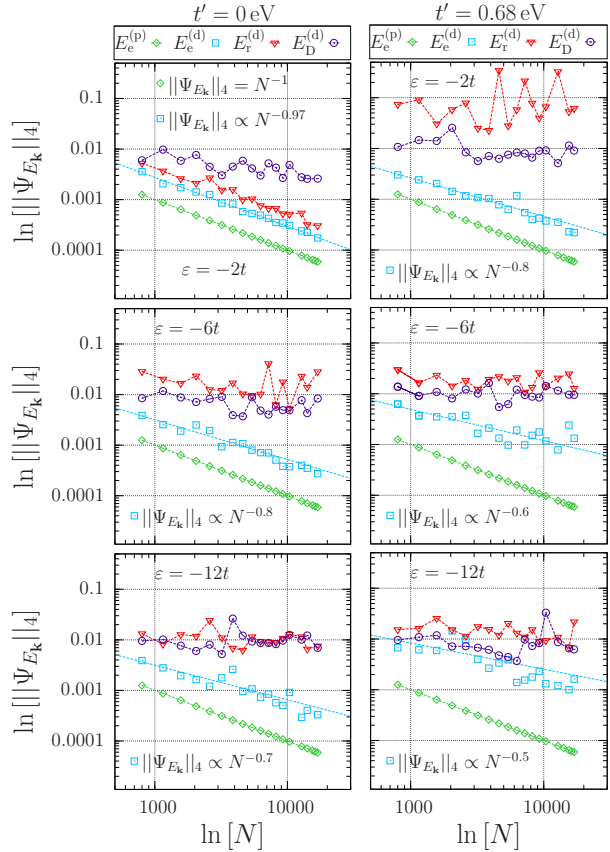}
\caption{(Color online) The $p=2$ IPR behavior as a function of $N$ for several selected energies, 
using different 
impurities self energies (from top to bottom) and without and with NNN interaction (left and right 
columns, respectively). The selected energies are the Dirac point energy [$E_{\rm D}$, (dark blue 
online) 
circles], the resonant state energy [$E_{\rm r}$, (red online) triangles] and $E_{\rm e}=E_{\rm D}-0.4t$ 
[(blue online) squares] for doped graphene $^{({\rm d})}$. In all cases, doped graphene has a $5\%$
impurity concentration. For the state with energy $E_e$, we present in all cases the  scaling exponent that results from the fitting, shown in the figures with (blue online) lines. For comparison proposes, we include 
the case of pure graphene for the energy $E_{\rm e}^{({\rm p})}$ [(green online) diamonds].}
\label{FigIPR2}
\end{figure}

To evaluate Eq. \eqref{iprs}, we calculated all the eigenvalues and eigenvectors of 
$\mathcal{H}$ by numerical diagonalization. The $p=2$ IPR behavior is shown in 
Figure~\ref{FigIPR2} as a function of $N$ for several selected energies, using different 
impurities self energies without and with NNN interaction. 
For the energy $E_{\rm e}=E_{\rm D}-0.4t$, far from the Dirac energy, we compare the behavior 
for pure graphene [(green online) diamonds] with doped graphene [(blue online) squares]. 
For pure graphene, the state is extended since the p=2 IPR goes like $N^{-1}$, as shown in 
Figure~\ref{FigIPR2}. For doped 
graphene, is clear that the  p=2 IPR can be fitted with a line, which suggests a non-localized, power 
law behavior. On the other hand, for states near the Dirac energy, in doped graphene the (dark blue 
online) circles and (red online) triangles behavior suggests localized states, since the p=2 IPR scales 
as $\propto N^0$. Thus, from this scaling analysis is clear that doped graphene, even in the absence 
of NNN interaction, presents different kinds of localization, 
as has been suggested in reference \cite{naumis} due to frustration effects \cite{barrios}, as 
well as in experiments \cite{bostwick}.

\begin{figure}[h]
\centering
\includegraphics[width=1.0\linewidth]{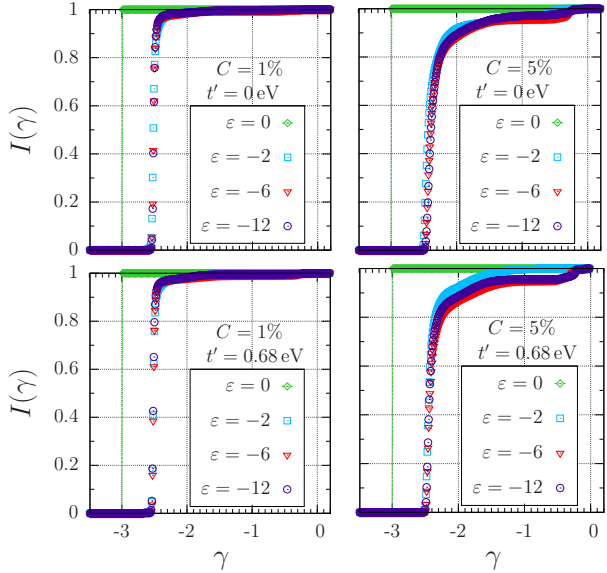}
\caption{(Color online) Integrated distribution of exponents for different impurity configurations using 
lattices with $N=16,928$ sites. Extended states are at $\gamma=-3$ and localized 
at $\gamma=0$. We observe that for pure graphene [(green online) diamonds], the distribution is a 
step function at $\gamma=-3$, 
meanwhile for doped graphene the distribution is shifted and it is no longer a step function.}
\label{FigIPR4}
\end{figure}

In order to know the exponent distribution of the power law behavior, we introduce 
the integrated distribution of exponents \cite{tsunetsugu},
\begin{align}
I(\gamma)=\frac{1}{N} \sum_{E_{\mathbf{k}}} \Theta \bigg(\gamma - 
\log_N \left[ ||\Psi_{E_{\mathbf{k}}}||_8\right] \bigg) \,,
\end{align}
where $\Theta$ is the step function. The exponent distribution is plotted in Figure~\ref{FigIPR4} for 
several impurity types and two different concentrations, $1\%$ and $5\%$. For pure graphene, all the 
states have the same scaling behavior, and a step is observed at  $\gamma=-3$. This means that all 
states have the same scaling, and thus all are extended, as expected from Bloch's theorem. However, 
for doped graphene, we observe two main effects.  First there is a shift to higher values of $\gamma$ 
and second, the jump is not anymore a discontinuity. Instead, we observe states that have a 
distribution of $\gamma$ values. In all cases, we observed that the minimal value of $\gamma$ is 
$\approx -5/2$, which means that the most extended states, follow a power law that goes as 
$r^{-3/8}$. States with $\gamma \approx 0$ are exponentially localized. Since no clear jump is 
observed in the  values of $\gamma$, it seems that there is a range of values for the exponents of the 
critical wave functions. Also, it is worthwhile mentioning that the presence of the NNN interaction 
preserves this behavior, which allows its experimental verification since the NNN is always present.

\begin{figure}[h]
\centering
\includegraphics[width=1.0\linewidth]{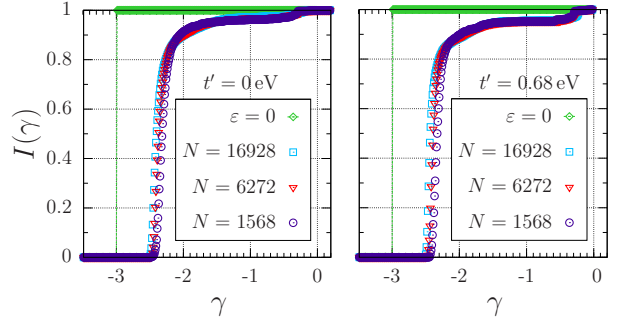}
\caption{(Color online) Example of the Integrated distribution of exponents for different sample sizes 
using a fixed concentration of impurities (5\%) with self-energy $\varepsilon=-6t$. 
The left panel corresponds 
to the model without NNN interaction, and the right panel graph with NNN.}
\label{FigIPR4Nx}
\end{figure}

To verify that such behavior is preserved as the system grows, in Fig.~\ref{FigIPR4Nx} we present the distribution $I(\gamma)$ for different  sample sizes. We can observe that the behavior is very similar at 
all sizes. From this analysis, we can conclude that there are many non-exponentially localized states, and that the power law behavior is not a finite size lattice effect. From the values of $\gamma$, these states are critical and non-normalizable. Finally, a careful check of such states reveals that 
localized states are near the Dirac point and at the band edges, while the power law non-normalizable states are near the middle part of the valence and conduction band, in agreement with previous theoretical arguments 
\cite{naumis,barrios}.

In conclusion, using a scaling analysis of the participation ratio, we have shown that the presence of 
disorder in graphene do not localize exponentially all states; instead, some states are critical
with a distribution of exponents. This result is robust against the inclusion of NNN interactions, 
in which the chirality is not preserved. This result is not only important for graphene, but it 
makes a revival of an old discussion concerning the possibility of having anomalous quantum 
percolation in two 
dimensional systems \cite{meir,grest}.

\begin{acknowledgments}
We thank the DGAPA-UNAM project IN-1003310-3. J.E. Barrios-Vargas acknowledges the 
scholarship from CONACyT (Mexico).
\end{acknowledgments}

\bibliographystyle{apsrev4-1}
\bibliography{biblioNormas}

\end{document}